\documentclass[aps,prb,twocolumn,showpacs]{revtex4}
\usepackage{latexsym}
\usepackage{amssymb}
\usepackage{graphicx}
\begin{document}
\title{Spin Hall Effect in a Diffusive Rashba Two-dimensional Electron Gas}
\author{S. Y. Liu}
\author{X. L. Lei}
\affiliation{Department of Physics, Shanghai Jiaotong University, 1954
Huashan Road, Shanghai 200030, China}
\date{\today}
\begin{abstract}
A nonequilibrium Green's functions approach to spin-Hall effect is
developed in a diffusive two-dimensional electron system with
Rashba spin-orbit interaction. In the presence of long-range
impurities, the coupled quantum kinetic equations are solved
analytically in the self-consistent Born approximation. It is
shown that the intrinsic spin-Hall effect stems from the
dc-field-induced perturbation of the density of states. In
addition, there is an additional disorder-mediated process, which
involves the transition of nonequilibrium electrons between two
spin-orbit-coupled bands. It results in an additional
collision-independent spin-Hall conductivity and leads to the
vanishing of the total spin-Hall current even at nonzero
temperature.

\end{abstract}

\maketitle

The discovery of spin-Hall effect, namely, the nonvanishing of
spin current along the direction perpendicular to the dc electric
field, has attracted much recent interest in the spin-dependent
transport in two-dimensional (2D) semiconductors with the
spin-orbit (SO) coupling. The early studies on this issue have
been devoted to the spin-orbit interaction between electrons and
impurities,\cite{DP,HS} and therefore, the spin-Hall effect
exhibits the {\it extrinsic} character. More recently, the
scattering-independent {\it intrinsic} spin-Hall effect, which
entirely stems from the internal-field-induced spin-orbit
coupling, has been predicted in the hole-doped
semiconductors\cite{t1} and two-dimensional semiconductors with
Rashba\cite{t2} and Dresselhaus SO coupling.\cite{t3}

In clean two-dimensional electron systems (2DES) with Rashba SO
interaction, Sinova {\it et al.} have shown that the spin-Hall
conductivity, $\sigma_{sH}$, has a universal value $e/8\pi$ at
zero temperature.\cite{t2} Subsequently, a great deal of works has
been focused on the effect of disorder on the spin-Hall current.
When the two-dimensional systems are sufficiently dirty and the
Anderson localization is dominant, the spin-Hall conductivity is
found to be independent of the disorder.\cite{t4} In diffusion
regime, the studies have been carried out when the presence of
short-range disorders. It has been demonstrated in
Ref.\,\onlinecite{Burkov,Nomura,t6} that the collision broadening can
reduce the spin-Hall current and the ballistic value of
$\sigma_{sH}$ can only be held for weak disorder. However, the
further investigation revealed that the persistent spin-hall
conductivity should be completely suppressed. This conclusion has
been obtained by different methods, such as the Kubo
formula,\cite{t7,t9,t8} the Keldysh formalism,\cite{t88} and
spin-density method\cite{t10} {\it etc}. It is commonly accepted
that such complete cancellation of spin-Hall current is not due to
any symmetry\cite{Murakami} and closely relates to the isotropy of
the electron-impurity scattering.\cite{t8}

However, for the realistic 2D systems, it is well known that the
disorder collision being short-range is a crude approximation and
the electron-impurity scattering is always long-ranged. In
Ref.\,\onlinecite{t8}, by considering the effect of long-range disorders
on the spin-Hall conductivity in 2D electron systems with Rashba
spin-orbit coupling, the nonvanishing of spin-Hall effect has been
shown due to the anisotropy of the collision. This argument does
not take the full effect of collision anisotropy, however and
hence a more careful investigation should be performed.\cite{t8}
In this letter, we construct a nonequilibrium Green's functions
approach to the spin-Hall effect in the diffusive Rashba
two-dimensional electrons. When the presence of the long-range
impurity-electron scattering, the derived kinetic equations are
solved analytically. We clarify that the dc-field-induced
intrinsic spin-Hall current is completely cancelled by the
spin-Hall effect, stemming from the disorder-induced transition of
the perturbative electrons, and in result, the total
collision-independent spin-Hall effect also vanishes when the
presence of long-range disorders.

We consider a quasi-2D electron semiconductor in the $x-y$ plane
subjected to the Rashba SO interaction. The single-particle
Hamiltonian of system can be written as
\begin{equation}
\hat h = \frac{{\bf p}^2}{2 m}+ \alpha {\bf p} \cdot ({\bf n}  \times{\vec \sigma}  ),\label{EH}
\end{equation}
where $\alpha$ is the Rashba coupling constant, ${\vec \sigma}$ are the Pauli matrices,
$m$ is the electron effective mass, and ${\bf n}$ is the unit vector perpendicular to
the 2DES plane.
By the local spinor unitary transformation
$\hat U({\bf p})$\cite{t9}
\begin{equation}
\hat U ( {\bf p} ) = \frac{1}{\sqrt{2}} \left(
\begin{array}{cc}
    1 & 1\\
    i{\rm e}^{i \varphi_{\bf p}} & - i e^{i \varphi_{\bf p}}
  \end{array}
 \right),
\end{equation}
the Hamiltonian (\ref{EH}) can be diagonalized, $\hat{h'}({\bf
p})=\hat U^+({\bf p})\hat{h}({\bf p}) \hat U({\bf p})={\rm
diag}(\varepsilon_1(p),\varepsilon_2(p))$ with
$\varepsilon_{\mu}(p)={\bf p}^2/2m+(-1)^\mu\alpha p$ and
$\mu=1,2$.

The nonequilibrium Green's functions $\hat{\rm G}^{r,<}({\bf p}, \omega)$ for the 2DES with Rashba coupling
are defined in the general manner and become $2\times 2$ matrices in spin space.\cite{DYX}
 Obviously,
in the helicity basis, where the single Hamiltonian is diagonal,
the noninteracting Green's functions are also diagonal
\begin{equation}
\hat{{\rm G}}^r_0({\bf p},\omega)={\rm diag}\left (\frac
1{\omega-\varepsilon_1(p)+i\varepsilon} ,\frac
1{\omega-\varepsilon_2(p)+i\varepsilon}\right ),
\end{equation}
\begin{equation}
\hat{{\rm G}}^<_0 ({\bf p},\omega)=-2i n_{\rm F}(\omega) {\rm Im} \hat{\rm G}^r_0 ({\bf p},\omega).
\end{equation}
Here $n_{\rm F}(\omega)$ is the Fermi function.

In quasi-2D semiconductors, the electrons should experience
scattering by impurities. The previous studies have only treated
the short-range interaction between electrons and impurities and
the relaxation is described by an isotropic parameter $\tau$.
However, when the presence of two spin-orbit-coupled bands, such a
simplification loses many important relaxation times, namely, the
longitudinal and transverse times. In this letter, we study more
realistic long-range collision and distinguish these different
scattering times carefully.

In the spin basis, the electron-impurity scattering process is
described through the long-range potential $V({\bf p}-{\bf k})$.
By transforming to the helicity basis it becomes $\hat{T}({\bf
p},{\bf k})=\hat U^+({\bf p})V({\bf p}-{\bf k})\hat U({\bf k})$. To deal
with the electron-impurity scattering, we restrict ourselves to
the self-consistent Born approximation, which corresponds to
the vertex corrections in the scheme of Kubo formalism. Hence, the
self-energies can take the form
\begin{equation}
\hat{\Sigma}^{r,<}({\bf p},\omega)=n_i\sum_{{\bf k}}\hat T({\bf
p},{\bf k})\hat{\rm G}^{r,<}({\bf k},\omega)\hat T^+({\bf p},{\bf
k}),\label{SE}
\end{equation}
with the impurity density $n_i$.
Putting the matrix $\hat{U}$ into Eq.\,(\ref{SE}) gives
\begin{widetext}
\begin{equation}
\Sigma^{r,<}({\bf p},T,t)=\frac 12 n_i\sum_{{\bf k}}|V({\bf
p}-{\bf k})|^2\left \{ a_1 {\rm G}^{r,<}+a_2\sigma_x{\rm
G}^{r,<}\sigma_x+ia_3[\sigma_x,{\rm G}^{r,<}]\right \}.
\end{equation}
\end{widetext}
Here $a_i$($i=1,2,3$) are the factors associated with the
directions of the momenta, $a_1=1+\cos (\phi_{\bf p}-\phi_{\bf
k})$, $a_2=1-\cos (\phi_{\bf p}-\phi_{\bf k})$, $a_3=\sin
(\phi_{\bf p}-\phi_{\bf k})$.

In order to investigate the transport properties of 2D system
driven by a dc electric field ${\bf E}$ along $x$ axis, the
dominant task is to carry out the nonequilibrium less Green's
functions $\hat {\rm G}^{<}$. The measurable quantities such as
current and spin-current, are closely related to them. It is well
known that the electric current operator in the spin basis reads
${\bf J}=\sum_{{\bf p}}\hat{\Psi}^+{\bf j}\hat{\Psi}$, with the
one-particle matrix current operator ${\bf j}$ being ${\bf j}=e
\left [\frac{{\bf p}}{m}+\alpha ({\bf n}\times{\vec \sigma})\right
]$. The current operator of the $i$-direction spin can be defined
through the electric current operator: ${\bf J}^i=({\bf
J}\sigma_i+\sigma_i{\bf J})/4e$.\cite{RA} We find, in the helicity
basis, the operator of $z$-direction-spin current along $y$ axis
is nondiagonal. By taking the statistical average, the Hall
spin-current can be computed via
\begin{equation}
{J}^{z}_y=\sum_{{\bf p}}\frac {p_y}{2m}[{\hat \rho}_{12}\left
({\bf p})+{\hat \rho}_{21}({\bf p})\right ],\label{Jz}
\end{equation}
with the distribution functions ${\hat \rho}_{\alpha \beta}({\bf p
})=-i\int \frac{{\rm d}\omega}{2\pi} \hat {\rm G}_{\alpha \beta}^<({\bf
p},\omega)$($\alpha,\beta=1,2$). It can be clearly seen that the
presence of spin-Hall effect is due to the nonvanishing of the
${\hat \rho}_{12}$ and ${\hat \rho}_{21}$, viz. interband
polarizations.

We follow the standard procedures described in Ref.\,\onlinecite{JH} to
derive the kinetic equations.
We first construct the equations for the less Green's functions in the
spin basis, and then modify it by the local unitary transformation.
Finally, for steady and homogeneous systems, we arrive at the kinetic
equations in the helicity basis
\begin{eqnarray}
ie {\bf {\rm E}} \cdot \left ({\bf\nabla}_{\bf p} \hat{\rm G}^< +
\frac {i{\bf \nabla}_{\bf p}\phi_{\bf p}}{2}[\hat{\rm G}^<,\sigma_x]
\right )
-\alpha  p {B}^< \nonumber\\
=( \hat{\Sigma}^r
   \hat{\rm G}^< - \hat{\rm G}^< \hat{\Sigma}^a - \hat{\rm G}^r \hat{\Sigma}^< +
   \hat{\Sigma}^< \hat{\rm G}^a )\label{EK}
\end{eqnarray}
with ${B}^<=[ \hat{\rm G}^<, \sigma_z]_-$. In this formula, the
arguments of the Green's functions $({\bf p},\omega)$ are dropped
for shortness. In deriving the equations, the scalar potential
gauge is used and only the lowest order of gradient expansion is
taken into account. Noted that the second term in the bracket of
the left hand of Eq.\,(\ref{EK}) arises from the fact that the
transformation is local. As we will show below, it will lead to
the intrinsic spin-Hall conductivity. At the same time, when the
electron-impurity collision is chosen to be short-ranged, these
kinetic equations reduce to the formula obtained in
Ref.\,\onlinecite{t88} by means of the Keldysh formalism.

We are only concerned with the linear response of the system to dc
fields. Hence, the kinetic equations can be linearized with
respect to the external electric fields. We rewrite the less
Green's functions as $\hat {\rm G}^<=\hat{\rm G}^<_0+\hat{\rm
G}^<_1$, with $\hat {\rm G}^<_0$ and $\hat{\rm G}^<_1$ being in
the zero- and first-order of $E$, respectively. As a result, the
kinetic equations for the less Green's functions $\hat {\rm G}^<_1$
have the forms
\begin{eqnarray}
-\alpha p \hat{C}_1+
ie{\rm \bf E}\cdot  {\bf \nabla}_{\bf p}\hat{\rm G}_0^<-\frac{1}{2} e {\bf \rm E} \cdot{\bf \nabla}_{\bf p}\phi_{\bf p} \hat{D}_0=\nonumber\\
( \hat{\Sigma}^r
   \hat{\rm G}^< - \hat{\rm G}^< \hat{\Sigma}^a - \hat{\rm G}^r \hat{\Sigma}^< +
   \hat{\Sigma}^< \hat{\rm G}^a )_1,\label{EQOne}
\end{eqnarray}
where the matrices $\hat{C}_1$ and $\hat{D}_0$ are
\begin{equation}
\hat{C}_1=\left (
\begin{array}{cc}
0&-2 (\hat{\rm G}^<_1)_{12}\\
2 (\hat{\rm G}^<_1)_{21}&0\\
\end{array}
\right ),
\end{equation}
\begin{equation}
\hat{D}_0=\left (
\begin{array}{cc}
0&(\hat{\rm G}^<_0)_{11}-(\hat{\rm G}^<_0)_{22}\\
(\hat{\rm G}^<_0)_{22}-(\hat{\rm G}^<_0)_{11}&0
\end{array}
\right ).\label{EG}
\end{equation}
The subscript $1$ in the right-hand side of Eq.\,(\ref{EQOne})
stands for keeping this term in the first-order of $E$.

From Eq.\,(\ref{EQOne}) we can see that two driving terms enter
the kinetic equations: $ie{\rm\bf E}\cdot {\bf \nabla}_{\bf
p}\hat{\rm G}_0^<$ and $ - e {\rm\bf  E} \cdot{\bf \nabla}_{\bf
p}\phi_{\bf p} \hat{D}_0/2$. Hence we can assume that the
solutions of these equations are formed from two terms $\hat {\rm
G}_1^{(1)}$ and $\hat{\rm G}_1^{(2)}$. In fact, when the disorder
becomes short-ranged, these two parts of solutions connect with
the terms in Kubo formalism. The first part of solutions stems
from the driving term completely and corresponds to the bubble
diagram in the scheme of Kubo formula.\cite{t7,t9,t8} At the same
time, the another part of solutions, relating to the
electron-impurity scattering, should be understood as the vertex
correction. In this way, the distribution functions in the first
order of dc field, $\hat \rho_1(t)$, can be written as $\hat
{\rho}_1(t) =\hat{\rho}_1^{(1)}(t)+\hat{\rho}_1^{(2)}(t)$ and,
correspondingly, the spin-Hall effect comes from two different
interband-polarization processes.

The less Green's functions $\hat {\rm G}^{(1)}_1$, associated with
the matrix $\hat{D}_0$, relate to the function $n_{\rm
F}(\omega)$. It is a nondiagonal matrix with same elements. In
fact, $\hat {\rm G}^{(1)}_1$ originate from the nondiagonal
elements of linear response retarded Green's functions, $(\hat{\rm
G}_1^{(1)})_{12}=(\hat{\rm G}^{(1)}_1)_{21}=-2in_{\rm F}{\rm Im}
(\hat{\rm G}_1^r)_{12}$. The latter one, $(\hat{\rm G}_1^r)_{12}$,
obeys the Dyson equation
\begin{eqnarray}
2\alpha p ({{\hat {\rm G}}_1}^r)_{12}+\frac 1{2p} eE\sin\phi_{\bf
p} [({{\hat {\rm G}}_0}^r)_{11}-({{\hat {\rm G}}_0}^r)_{22}]
=\nonumber \\
({{\hat {\rm G}}_1}^r)_{12}[({{\hat \Sigma}_0}^r)_{11}-({{\hat \Sigma}_0}^r)_{22}]
-({{\hat \Sigma}_1}^r)_{12}[({{\hat {\rm G}}_0}^r)_{11}-({{\hat {\rm G}}_0}^r)_{22}],\label{EGR}
\end{eqnarray}
and can be obtained analytically as
\begin{equation}
({{\hat {\rm G}}_1}^r)_{12}=-\frac {eE}{4 \alpha p^2} \sin
\phi_{\bf p} [({{\hat {\rm G}}_0}^r)_{11}-({{\hat {\rm
G}}_0}^r)_{22}].\label{R1}
\end{equation}

It is well known that the linear response can not disturb the
retarded Green's functions in semiconductors without spin-orbit
interaction.\cite{M1} However, when the spin-orbit coupling is
introduced, even weak dc field can produce the transition between
the spin-orbit-coupled bands, resulting in interband polarization.
The factor $n_{\rm F}(\omega)$ in the less Green's functions
$\hat{\rm G}^{(1)}_1$ indicates that all electrons join in
this process.

Substituting the Eq.\,(\ref{R1}) into (\ref{Jz}) and neglecting
the broadening of the noninteracting retarded Green's functions,
we find the contribution from the first polarization
process to the spin-Hall conductivity
\begin{equation}
\sigma_{sH}^{(1)}=\frac{-e}{16\pi m\alpha}\int_0^{\infty}{\rm d} p
n_{\rm F}(\varepsilon_1(p)-\mu)-n_{\rm F}(\varepsilon_2(p)-\mu),
\end{equation}
where $\mu$ is the chemical potential. At zero temperature,
$\sigma_{sH}^{(1)}$ approaches the ballistic value $e/8\pi$. Note
that this formula agrees with that obtained in the previous
studies.\cite{t6,t9,t8}

The remaining part of Eq.\,(\ref{EQOne}) describes the transport
process and is associated with the function $\partial n_{\rm
F}(\omega)/\partial \omega$. To solve it, we employ the two-band
generalized Kadanoff-Baym ansatz (GKBA). This ansatz is widely
used in the quantum kinetics of semiconductors driven by ac
fields\cite{GKBA,GKBA1} and accurate enough to yield the
quantitative agreements with experiments.\cite{JH} To study the
linear response of 2DEG with SO coupling, we only need to consider
the GKBA in the first order of dc field strength $E$. To further
simplify the treatment, we ignore the broadening of $\hat{\rm
G}^{r,a}_0$ induced by electron-impurity scattering. It is well
known, this approximation combined with the gradient approximation
are referred as the Boltzmann limit and valid when the spatial and
temporal variations change slowly.\cite{JH} They are effective for
modelling the quasiclassical optical and transport processes in
semiconductors.

By taking into account the time-independence of $\hat \rho (t)$ in
the steady-state transport, the second part of solution of
Eq.\,(\ref{EQOne}) can be obtained analytically. Retaining the
results to ${\rm O}(n_i)$, we find that the functions $(\hat
\rho_1^{(2)})_{12}$ can be expressed as
\begin{equation}
(\hat \rho_1^{(2)}) _{12}({\bf p})=\zeta (p)eE\sin \phi_{\bf p} +i
\xi (p) eE \sin \phi_{\bf p},
\end{equation}
where $\zeta (p)$ and $\xi (p)$ are real functions. Only the
real part of $(\hat {\rho}_1^{(2)})_{12}$ needs to be calculated,
since the contribution of the imaginary part to the spin Hall
current vanishes due to the symmetry relation $(\hat
{\rho}_1^{(2)})_{12}= (\hat {\rho}_1^{*(2)})_{21}$.

The elementary calculation yields
\begin{widetext}
\begin{eqnarray}
\zeta (p) &=&\frac 1{4\alpha p} \left \{ \nabla_E n_{\rm F} (E)
\left |_{E=\varepsilon_2(p)}(\alpha+p/m)
-\nabla_E n_{\rm F} (E) \right |_{E=\varepsilon_1 (p)}(-\alpha+p/m)\right .\nonumber\\
&&\left .+\frac 1 {\tau_{221}}\left [\Theta_2
(p)+\Theta_1(p+2m\alpha)\right ](\alpha+p/m) -\frac 1
{\tau_{212}}\left [\Theta_1 (p)+\Theta_2(p-2m\alpha)\right
](-\alpha+p/m) \right \}.\label{ZZZ}
\end{eqnarray}
\end{widetext}
The functions $\Theta_1(p)$ and $\Theta_2(p)$, coupled by the
equations,
\begin{equation}
-\nabla_E n_{\rm F} (E) |_{E=\varepsilon_\mu(p)}=
\frac{\Theta_\mu(p)}{\tau_{1\mu\mu}} +\frac{\Theta_\mu
(p)}{\tau_{2\mu\bar{\mu}}} -\frac{\Theta_{\bar{\mu}} (p+(-1)^\mu
2m \alpha)}{\tau_{3\mu\bar{\mu}}},\label{EQR1}
\end{equation}
connect to the diagonal distribution functions
\begin{equation}
(\hat \rho_1) _{\mu\mu} ({\bf p})=[(-1)^\mu \alpha
+p/m]\Theta_{\mu}(p)eE\cos \phi_{\bf p},
\end{equation}
with $\mu=1,2$ and $\bar{\mu}=3-\mu$. In these equations, several
different relaxation times, defined by
\begin{equation}
\frac {1}{\tau_{i\mu\nu}}=2\pi n_i\sum_k |V({\bf p}-{\bf k})|^2
\Lambda_{i\mu\nu} (\phi_{{\bf k} -{\bf p}},p,k),
\end{equation}
emerge due to the long-range potential. Here
the $\Lambda_{i\mu\nu}$ are
$\Lambda_{1\mu\nu}(\phi,p,k)=\frac 12 \sin ^2 \phi \delta
(\varepsilon _{\mu p}-\varepsilon _{\nu k})$,
$\Lambda_{2\mu\nu}(\phi,p,k)=\frac 12 (1-\cos \phi) \delta
(\varepsilon _{\mu p}-\varepsilon _{\nu k})$ and
$\Lambda_{3\mu\nu}(\phi,p,k)=\frac 12 \cos \phi(1-\cos \phi)
\delta (\varepsilon _{\mu p}-\varepsilon _{\nu k})$.

The contribution of second interband polarization process to the
spin-Hall conductivity can be evaluated by
\begin{equation}
\sigma_{sH}^{(2)}=\frac e{4\pi}\int_0^{\infty}\frac {p^2}m
\zeta(p){\rm d}p.
\end{equation}
It should be noted that the spin-Hall conductivity coming from the
terms in the second line of Eq.\,(\ref{ZZZ}) is zero at zero temperature.
This is straightforwardly educed from the fact that zero-temperature
difference between the Fermi momenta for the two
spin-orbit-coupled bands is $2m\alpha$ and the factor $\nabla_E
n_{\rm F}(E)$ reduces to the delta function.
Omitting these terms at nonzero temperature is also supported
by numerical estimation. At temperature lower than $4.2$\,K,
considering long-range collision between the remote impurities and
electrons in GaAs/AlGaAs heterojunctions,\cite{RMP} we find the
magnitude of contributions of these terms to the spin-Hall
conductivity more than five orders smaller than that from the
other terms. Hence, the spin-Hall conductivity $\sigma_{sH}^{(2)}$
can be simplified as
\begin{eqnarray}
\sigma_{sH}^{(2)}&=&\frac e{16\pi m\alpha}\int_0^{\infty}p {\rm
d}p \left \{ \nabla_E n_{\rm F} (E) \left |_{E=\varepsilon_2(p)}
(\alpha+p/m)
\right .\right .\nonumber\\
&&\left .\left .-\nabla_E n_{\rm F} (E) \right
|_{E=\varepsilon_1(p)}(-\alpha+p/m)\right \}.
\end{eqnarray}
Finally, we obtain $\sigma_{sH}^{(2)}=-\sigma_{sH}^{(1)}$.

From the Eqs.\,(\ref{ZZZ}) and (\ref{EQR1}) we can see that the
spin-Hall conductivity $\sigma_{sH}^{(2)}$ appears from the
transport process. In the dilute impurity limit, the diagonal
distribution functions of nonequilibrium electrons driven by dc
fields are of order of $n_i^{-1}$. These perturbed electrons in
one band can transit to another band, resulting in interband
polarization. The latter process is proportional to $n_i$ and,
therefore, the nondiagonal distribution functions in the lowest
order of $n_i$ appear. Hence, it is obvious that, here, the
disorder plays an intermediate role. Such disorder-mediated
process, in contrast to the one produced by all electrons,
contributes from the electrons near the Fermi surfaces.

Note that in our treatment the influence of broadening of Green's
functions on the spin-Hall effect remains untouched. This implies
that our results are useful for the relative clean samples. At the
same time, the derived kinetic equations in this letter can serve
as the start point for further studying the effect of collision
broadening, which is expected to affect on the linear response of
the retarded and less Green's functions in different fashions.
Hence, the spin-Hall effect may be observed in dirty samples.

In conclusion, we have presented a nonequilibrium Green's
functions approach to the spin-Hall effect in Rashba
two-dimensional electron systems. By taking into account the
long-range scattering between impurities and electrons, the
intrinsic spin-Hall conductivity has been recovered. It has also
been illustrated that the interband transition of nonequilibrium
electrons, participating in longitudinal transport, results in a
disorder-mediated spin-Hall effect. As the result, the total
spin-Hall effect disappears.

We would like to thank Drs. S.-Q. Shen and J. Sinova for useful
information. One of the authors (SYL) gratefully acknowledges
invaluable discussions with Drs. W. S. Liu, Y. Chen and M. W. Wu.
This work was supported by the National Science Foundation of
China, the Special Funds for Major State Basic
 Research Project, and the Youth
Scientific Research Startup Foundation of SJTU.

\end{document}